\documentclass{osa-article}
\journal{oe}

\DeclareMathAlphabet\mathbfcal{OMS}{cmsy}{b}{n}

\begin{document}

\title{Optimization of intensity-modulation/direct-detection optical key distribution under passive eavesdropping}

\author{
Konrad Banaszek,\authormark{1,2,*} Micha{\l} Jachura,\authormark{1} Piotr Kolenderski,\authormark{3} and Miko{\l}aj Lasota\authormark{3}}

\address{\authormark{1}Centre for Quantum Optical Technologies, Centre of New Technologies,
University of Warsaw, Banacha 2c, 02-097 Warszawa, Poland}
\address{\authormark{2}Faculty of Physics, University of Warsaw, Pasteura 5, 02-093 Warszawa, Poland}
\address{\authormark{3}Faculty of Physics, Astronomy and Informatics, Nicolaus Copernicus
University, Grudzi\k{a}dzka 5, 87-100 Toru\'{n}, Poland
}
\email{\authormark{*}k.banaszek@uw.edu.pl} %% email address is required



\begin{abstract}
We analyze theoretically optimal operation of an optical key distribution (OKD) link based on fine intensity modulation of an optical signal transmitted over an attenuating channel to a direct detection receiver. With suitable digital postprocessing, the users may generate a secret key that will be unknown to an unauthorized party collecting passively a fraction of the signal that escapes detection by the legitimate recipient. The security is ensured by the presence of the shot noise that inevitably accompanies eavesdropper's attempt to detect the collected signal. It is shown that the key amount depends on a ratio that compares legitimate recipient's and eavesdropper's capabilities to detect the signal, including noise contributed by their respective detectors. A simple proportionality relation is derived in the strong eavesdropping regime and closed expressions for the optimal depth of binary intensity modulation as well as the discrimination thresholds for hard-decoded direct detection are given. The presented results substantially simplify design of practical OKD systems operating under changing external conditions, e.g.\ variable atmospheric absorption in the case of free-space optical links.
\end{abstract}

%%%%%%%%%%%%%%%%%%%%%%%%%%  Body  %%%%%%%%%%%%%%%%%%%%%%%%%%

%-----------------------------------------------------------------------------------------

\section{Introduction}

Modern day cyber security relies primarily on software solutions. However, there is a growing recognition for the need to protect also  the physical layer of communication systems \cite{BlochBarros2011}. One of the emerging techniques in this field is quantum key distribution (QKD) \cite{ScaraniRMP2009,XuRMP2020,PirandolaAOP2020}, which allows two parties, customarily called Alice and Bob, to generate a secure key. Any eavesdropping attack carried out by an unauthorised third party, usually referred to as Eve, can be detected by legitimate users as errors in the generated raw key. Such an attack reduces the amount of a secure key after the privacy amplification step or, if forceful enough, renders its generation impossible. The ambition of QKD technology is to make the key distribution secure even against the most sophisticated physical attacks permitted by quantum theory, including use of a quantum processor to manipulate quantum signals transmitted between Alice and Bob, or exploitation of side channels \cite{Derkach2016,Derkach2017,Pereira2018,Jain2021}. However, this ambitious goal is inseparably entwined with stringent requirements on noise properties of components used in QKD systems and the quality of the communication channel. In practice, such imperfections markedly limit the range of QKD links and the attainable key rates.

An interesting alternative explored in recent years is to assume a restricted class of eavesdropping attacks that can be viewed as imminent with current or near-term technology. One relevant scenario is to consider the actual threat in the form of passive eavesdropping, i.e.\ Eve's ability to access a fraction of the signal that does not reach Bob's receiver. Such a threat is formally equivalent to the beam splitting attack considered in the security analysis of QKD systems \cite{Pan2020}. As recently proposed \cite{IkutaNJP2016} and discussed in a number of following works \cite{TrinhIEEEAccess2018,ErikssonOE2018,YamamoriJJAP2020,TrinhIEEETC2020}, in such a scenario the key security can be ensured by shot noise properties of the electromagnetic radiation. In order to distinguish this approach from QKD protocols, which rely essentially on incompatible quantum measurements, we will use for the former the designation of {\em optical key distribution (OKD)}, as the optical band is the most obvious choice for communication links operated at or near the shot noise limit.

The purpose of this paper is to analyze theoretically optimization of an OKD link based on intensity modulation/direct detection (IM/DD) transmission, which is one of the simplest options for communication systems. It is shown that when the photodetection noise is modelled using Gaussian statistics, the attainable key rate depends on a simple quantity that compares Bob's and Eve's capabilities to detect the signal sent by Alice, including noise contributed by their respective detection subsystems. Furthermore, in the strong eavesdropping regime, when Eve's capability significantly exceeds that of Bob, we give simple recipes for determining the optimal signal modulation depth used in Alice's transmitter, as well as discrimination thresholds implemented in Bob's receiver. This regime is especially relevant to long-haul free-space optical (FSO) links, where Bob's device can often capture only a small fraction of the optical signal produced by Alice, while a substantially higher portion may be available to Eve.
General results presented here substantially simplify realistic modelling of practical OKD links, that needs to include e.g.\ variable atmospheric conditions.
The approach to physical layer security explored here can be viewed as complementary to encodings designed for quantum wiretap channels \cite{EndoIEEEPhJ2015,FujiwaraOpEx2018,PanPRApplied2020,VazquezCastroPRApplied2021}, which usually give the legitimate user an advantage in collecting the optical signal compared to an eavesdropper. In contrast, postselection of detection events used in OKD enables Alice and Bob to generate a secure key even when Eve has access to a substantially larger fraction of the signal compared to Bob.

This work is organized as follows. Sec.~\ref{Sec:Model} describes the model of an OKD link investigated in subsequent parts of the paper. Sec.~\ref{Sec:Gaussian} analyzes the case of Gaussian signal modulation, which will serve later as a reference for the simpler and more practical technique of binary modulation. The latter is discussed in Sec.~\ref{Sec:Binary} considering both soft and hard decoding at Bob's receiver. Finally, Sec.~\ref{Sec:Conclusions} concludes the paper.

\section{Model}
\label{Sec:Model}

\begin{figure}
\centering
\includegraphics[width=0.75\columnwidth]{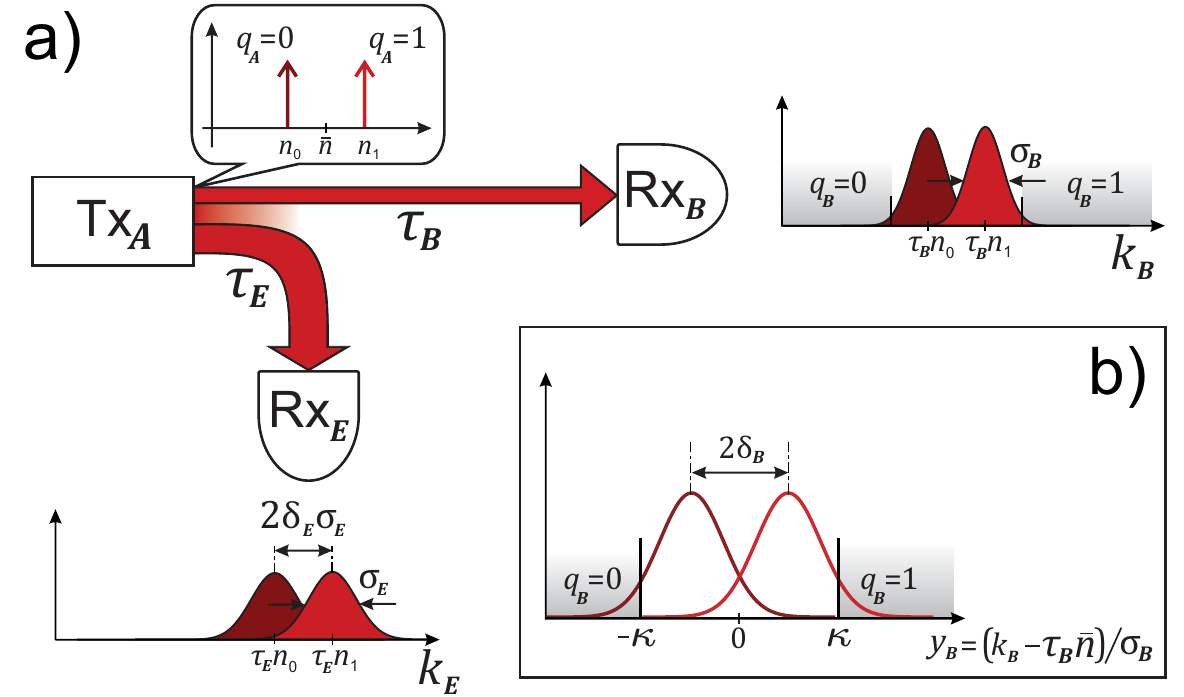}
\caption{(a) Optical key distribution under passive eavesdropping. Alice's transmitter $\textrm{Tx}_A$ generates a finely intensity-modulated signal with mean optical energy per slot $\bar{n}$. In the simplest case of binary modulation shown in the inset, one of two optical energies $n_0$ or $n_1$ is chosen according to Alice's bit value $q_A=0,1$. Bob's and Eve's receivers $\textrm{Rx}_B$ and $\textrm{Rx}_E$ detect respectively fractions $\tau_B$ and $\tau_E$ of the signal. The model assumes Gaussian detection noise with respective variances $\sigma_B^2$ and $\sigma_E^2$. A convenient figure to characterize the binary modulation depth is $\delta_E = \tau_E(n_1-n_0)/(2\sigma_E)$. Binary-modulated signal can be discriminated using two thresholds located symmetrically around the mean value, as depicted at Bob's location. (b) Statistics of Bob's outcomes for binary-modulated signal represented using a normalized variable $y_B=(k_B - \tau_B \bar{n})/ \sigma_B$. Discrimination thresholds are set at $\pm \kappa$.}
\label{Fig:Scheme}
\end{figure}

As shown in Fig.~\ref{Fig:Scheme}, Alice generates the optical signal by modulating finely the intensity of laser light carried in discrete temporal slots. In order to facilitate the analysis of shot noise effects it will be convenient to specify the emitted optical energy as the mean number $n_A$ of photons contained in an individual slot. The average optical energy per slot will be denoted as $\mathbb{E}[n_A] = \bar{n}$. A fraction $\tau_B$ of the emitted signal reaches Bob's receiver, whereas a fraction $\tau_E$ can be accessed by Eve. Bob measures the intensity of the received signal by means of direct detection.
%, which at the fundamental level is limited by shot noise.
If the phase of the optical field is random between individual slots, Eve is also left with the intensity measurement to determine the signal modulation. It will be assumed that the optical signals received by Bob and Eve are strong enough to justify Gaussian statistics for the detection outcomes, denoted respectively by variables $k_B$ and $k_E$. In the case of small modulation depth it is reasonable to assume that the variances $\sigma_B^2$ and $\sigma_E^2$ of Bob's and Eve's detectors are independent of the mean photon number received in a given slot. Thus the conditional distributions for Bob's and Eve's outcomes for a given optical energy $n_A$ sent by Alice will be:
\begin{equation}
k_B | n_A \sim {\cal N}(\tau_B n_A, \sigma_B^2), \qquad k_E | n_A \sim {\cal N} (\tau_E n_A, \sigma_E^2).
\end{equation}
For shot noise limited detection of a weakly modulated signal one can take to a good approximation $\sigma_B^2= \tau_B \bar{n}$ and $\sigma_E^2= \tau_E \bar{n}$ \cite{BanaszekJLT2020}.

The most intuitive protocol for key distribution in the setup described above is based on binary modulation \cite{IkutaNJP2016}, when the optical energy $n_A$ in each temporal slot assumes one of two equiprobable values $n_0$ or $n_1$ satisfying $(n_0+n_1)/2 = \bar{n}$ that are selected according to Alice's random bit value $q_A=0,1$. For concreteness, let us take $n_1 > n_0$. Suppose that Bob selects two discrimination thresholds located symmetrically around $\tau_B \bar{n}$. For sufficiently large spacing between the thresholds, obtaining a detection event in one of the two outer regions enables Bob to identify almost unambiguously the value of $q_A$. Sifted slots for which such events have been obtained are communicated by Bob to Alice over a public channel. The recorded bit values form the raw key between Alice and Bob.
If the modulation depth is so low that the distributions of Eve's outcomes $k_E$  for $n_A = n_0$ and $n_A = n_1$ overlap substantially, i.e.\ $\tau_E(n_1 - n_0) \lesssim \sigma_E$, Eve cannot learn much about the sifted key. For example, if she implements dual-threshold discrimination analogous to that used by Bob, most key generating events will produce outcomes $k_E$ that lie in the central region. In general it is advantageous for Eve to gain information about the key from the ``soft'' values $k_E$, without threshold discrimination \cite{ErikssonOE2018}. This more powerful eavesdropping strategy will be investigated in the present work.

The basic theoretical tool in the security analysis will be the Csisz\'{a}r-K\"{o}rner formula \cite{CsiszarKornerIEEETIT1978}:
\begin{equation}
\label{Eq:KeyGeneral}
{\sf K} = \max \{{\sf I}(A;B) - {\sf I}(B;E), 0 \}.
\end{equation}
which expresses the attainable key per slot ${\sf K}$ as the mutual information ${\sf I}(A;B)$ between the legitimate users Alice and Bob reduced by the mutual information ${\sf I}(B;E)$ specifying how much Eve might have learnt about Bob's outcomes. The above expression describes the reverse reconciliation scenario, when Alice corrects her raw key according to information received from Bob. This scenario is more robust against eavesdropping compared to the direct reconciliation approach, in which Alice's and Bob's roles in correcting the raw key are reversed. One can note here a formal similarity with reconciliation in continuous-variable QKD protocols \cite{SilberhornPRL2002,GrosshansNAT2003,GrosshansQIC2003,Usenko2011,LaudenbachAQT2018}.

\section{Gaussian modulation}
\label{Sec:Gaussian}

As a reference case that can be solved in a closed analytical form, we will consider first Gaussian modulation implemented by Alice, where the mean photon number in an individual temporal slot follows a normal distribution with a variance $\sigma_{A}^2$:
\begin{equation}
n_A \sim {\cal N} (\bar{n}, \sigma_A^2 ).
\end{equation}
In the following, it will be convenient to use shifted and normalized variables defined by
\begin{equation}
y_A = (n_A-\bar{n})/\sigma_A  \sim {\cal N} (0,1)
\end{equation}
for Alice and
\begin{equation}
y_B =  (k_B-\tau_B\bar{n})/ \sigma_B, \qquad
y_E =  (k_E-\tau_E\bar{n}) / \sigma_E
\end{equation}
for Bob and Eve, respectively. In the Gaussian scenario, Bob and Eve process ``soft'' outcomes $y_B$ and $y_B$ without threshold discrimination. Therefore mutual information in Eq.~(\ref{Eq:KeyGeneral}) needs to be calculated for pairs of Gaussian variables $(y_A,y_B)$ and $(y_B, y_E)$.

\begin{figure}
\centering
\includegraphics[width=0.6\columnwidth]{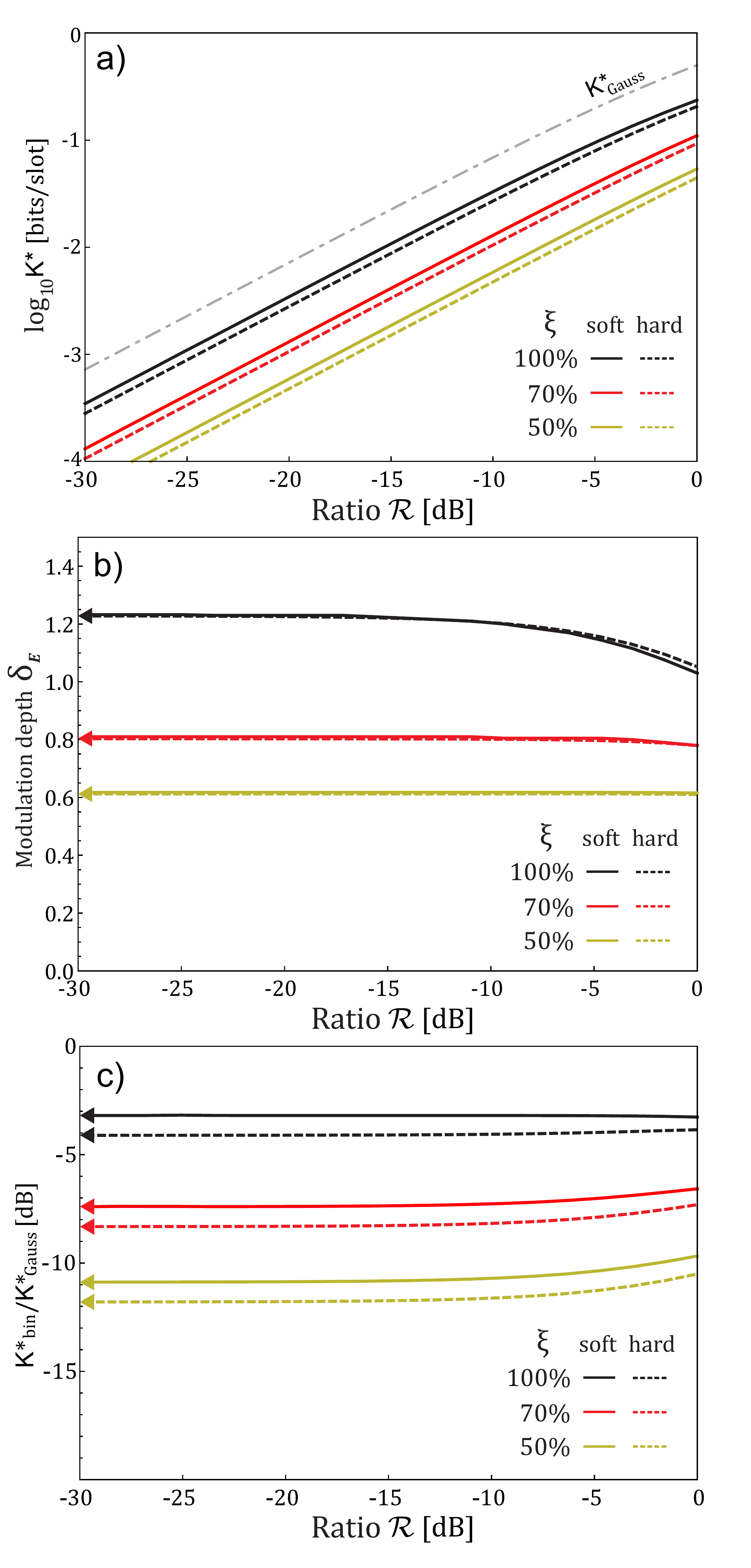}
\caption{(a) The  key limit for Gaussian modulation ${\sf K}_{\textrm{Gauss}}^\ast$ (grey dashed-dotted line) and the binary modulation assuming soft ${\sf K}^\ast_{{\textrm{bin}},{\textrm{soft}}}$ (solid lines) and hard
${\sf K}^\ast_{{\textrm{bin}},{\textrm{hard}}}$ (dashed lines) decoding, plotted as a function of the ratio ${\cal R}$ for the reconciliation efficiencies $\xi =100\%$ (black), $70\%$ (red), and $50\%$ (gold). (b) Corresponding optimal binary modulation depth, characterized by $\delta_E = \tau_E(n_1-n_0)/\sigma_E$. (c) The key reduction compared to the Gaussian limit ${\sf K}^\ast_{{\textrm{bin}},{\textrm{soft}}}/{\sf K}_{\textrm{Gauss}}^\ast$ (solid lines) and ${\sf K}^\ast_{{\textrm{bin}},{\textrm{hard}}}/{\sf K}_{\textrm{Gauss}}^\ast$ (dashed lines). The asymptotic values in the limit ${\cal R} \rightarrow 0$ based on Eqs.~(\ref{Eq:K(binsoft)approx}), (\ref{Eq:gamma(beta)def}) and (\ref{Eq:Keybinhardapprox}) are indicated with arrows on the vertical axes of panels (b,c).}
\label{Fig:Key}
\end{figure}

It is straightforward to obtain that the covariance matrix $\mathbf{C}$ for the triplet of the random variables $(y_A, y_B, y_E)$ reads
\begin{equation}
\mathbf{C} =
\begin{pmatrix}
1   & \tau_B\sigma_A /\sigma_B & \tau_E \sigma_A /\sigma_E\\
\tau_B \sigma_A/\sigma_B  & 1 + (\tau_B \sigma_A/\sigma_B)^2&
(\tau_B\sigma_A /\sigma_B)(\tau_E \sigma_A /\sigma_E) \\
\tau_E \sigma_A /\sigma_E &
(\tau_B\sigma_A /\sigma_B)(\tau_E \sigma_A /\sigma_E)
& 1 + (\tau_E \sigma_A /\sigma_E)^2
\end{pmatrix}.
\end{equation}
Applying the standard expression for the mutual information for a pair of Gaussian variables $(y_\mu, y_\nu)$ given by
\begin{equation}
{\sf I}(\mu;\nu) = - {\textstyle\frac{1}{2}} \log_2 \left( 1 -
\frac{\bigl( \textrm{Cov}[y_\mu,y_\nu]\bigr)^2}{\textrm{Var}[y_\mu]\textrm{Var}[y_\nu]}\right),
\end{equation}
yields for Alice and Bob
\begin{equation}
{\sf I}(A;B) =  {\textstyle\frac{1}{2}} \log_2 [ 1 + (\tau_B \sigma_A/\sigma_B)^2],
\end{equation}
whereas for Bob and Eve one has
\begin{equation}
{\sf I}(B;E) = {\textstyle\frac{1}{2}} \log_2 \left( \frac{[1 + (\tau_B \sigma_A/\sigma_B)^2]
[1 + (\tau_E \sigma_A/\sigma_E)^2]}{1 + (\tau_B \sigma_A/\sigma_B)^2
+ (\tau_E \sigma_A/\sigma_E)^2} \right).
\end{equation}
The expression for the key in the reverse reconcilliation scenario calculated according to Eq.~(\ref{Eq:KeyGeneral}) can be simplified to the form
\begin{equation}
{\sf K}_{\textrm{Gauss}} = {\sf I}(A;B) - {\sf I}(B;E) =
{\textstyle\frac{1}{2}} \log_2 \left( 1 + \frac{(\tau_B\sigma_A/\sigma_B)^2}{1+ (\tau_E \sigma_A /\sigma_E)^2}\right).
\end{equation}
It will be convenient to denote $ s= \tau_E \sigma_A/\sigma_E$ and write
\begin{equation}
\label{Eq:KG}
{\sf K}_{\textrm{Gauss}} = {\textstyle\frac{1}{2}} \log_2 \left( 1 +  {\cal R} \frac{s^2}{1+s^2}\right)
\end{equation}
where the ratio ${\cal R}$, defined by
\begin{equation}
\label{Eq:Rdef}
{\cal R} = \left(\frac{\tau_B \sigma_E}{\tau_E\sigma_B}\right)^2,
\end{equation}
compares Bob's capability to detect the signal to that of Eve. It is immediately seen that
the right-hand side of Eq.~(\ref{Eq:KG}) is a monotonically increasing function of $s$ and the maximum is reached asymptotically for $s \rightarrow \infty$,
\begin{equation}
\label{Eq:KGasympt}
{\sf K}_{\textrm{Gauss}} \le {\sf K}_{\textrm{Gauss}}^\ast = {\textstyle\frac{1}{2}} \log_2 ( 1 +  {\cal R} ) \approx
{\textstyle\frac{1}{2}} {\cal R} \log_2 e,
\end{equation}
where the last, approximate form of the upper bound on the key is valid when ${\cal R} \ll 1$, i.e.\ when Eve has much stronger capability to detect the signal compared to Bob. This regime will be designated as the strong eavesdropping scenario and will be the main focus of further analysis. The upper key bound ${\sf K}_{\textrm{Gauss}}^\ast$ is plotted as a function of ${\cal R}$ in Fig.~\ref{Fig:Key}(a).

From the perspective of legitimate users it is natural to assume the worst-case scenario that Eve's detector operates at the shot noise level, i.e.\ $\sigma_E^2 = \tau_E \bar{n}$. Taking Bob's detector variance as a sum of contributions produced by the shot noise and the detector thermal noise, $\sigma_B^2 = \tau_B \bar{n} + \sigma_{B, {\textrm{th}}}^2$, yields:
\begin{equation}
{\cal R} = \frac{\tau_B}{\tau_E} \left(1 + \frac{\sigma_{B, {\textrm{th}}}^2}{\tau_B \bar{n}}\right)^{-1}
\end{equation}
The second term in the round brackets can be interpreted as Bob's detector thermal noise variance specified in shot noise units. When Bob's detector operates at the shot noise level, the above expression reduces to a simple ratio of channel transmissions to Bob and to Eve, ${\cal R} = \tau_B/\tau_E$.

Note that although the upper key bound ${\sf K}_{\textrm{Gauss}}^\ast$ given in (\ref{Eq:KGasympt}) is attained in the asymptotic limit of an infinite variance of Alice modulation, the key amount becomes comparable with ${\sf K}_{\textrm{Gauss}}^\ast$ when $s \gtrsim 1$, which translates into the condition $\tau_E \sigma_A \gtrsim \sigma_E$. This means that fluctuations of Eve's detection outcome introduced by Alice's modulation should be at least as large as the noise of Eve's detection.
%%%This condition can be easily satisfied while staying in the weak modulation depth regime.

\section{Binary modulation}
\label{Sec:Binary}

In a more practical scenario, Alice uses binary modulation and choses the optical energy $n_{q_A}$ determined by the value of her bit $q_A=0,1$, as described in Sec.~\ref{Sec:Model}.
The average optical energy constraint implies that $(n_0+n_1)/2= \bar{n}$. The normalized variables describing outcomes of Bob's and Eve's detection are given by:
\begin{equation}
y_B | q_A \sim {\cal N} \bigl( (-1)^{q_A +1} \delta_B , 1\bigr), \qquad
y_E | q_A \sim {\cal N} \bigl( (-1)^{q_A +1} \delta_E , 1\bigr)
\end{equation}
Here the parameters
\begin{equation}
\delta_B = \tau_B (n_1-n_0)/(2\sigma_B), \qquad
\delta_E = \tau_E (n_1-n_0) / (2\sigma_E)
\end{equation}
characterize the modulation depth as detected respectively by Bob's and Eve's receivers and mapped onto normalized variables $y_B$ and $y_E$. Note that these two parameters are related through a simple rescaling:
\begin{equation}
\label{Eq:deltaBdeltaErel}
\delta_B = \sqrt{\cal R} \delta_E,
\end{equation}
where the parameter ${\cal R}$ has been defined in Eq.~(\ref{Eq:Rdef}).
In order to make the discussion more realistic, the expression for the key given in Eq.~(\ref{Eq:KeyGeneral}) will be replaced by a more general formula
\begin{equation}
\label{Eq:Keybeta}
{\sf K}_{\textrm{bin}} = \xi{\sf I}(A;B) - {\sf I}(B;E)
\end{equation}
where the positive factor $\xi$ characterizes reconciliation efficiency that is usually lower than its Shannon limit, $\xi \le 1$. Further analysis depends on whether Bob extracts the key from the continuous outcomes $y_B$ themselves (soft decoding), or applies dual-threshold discrimination to convert in a fraction of cases the outcome $y_B$ into a binary variable $q_B$ (hard decoding), as described qualitatively in Sec.~\ref{Sec:Model}.

\subsection{Soft decoding}

\begin{figure}
\centering
\includegraphics[width=1\columnwidth]{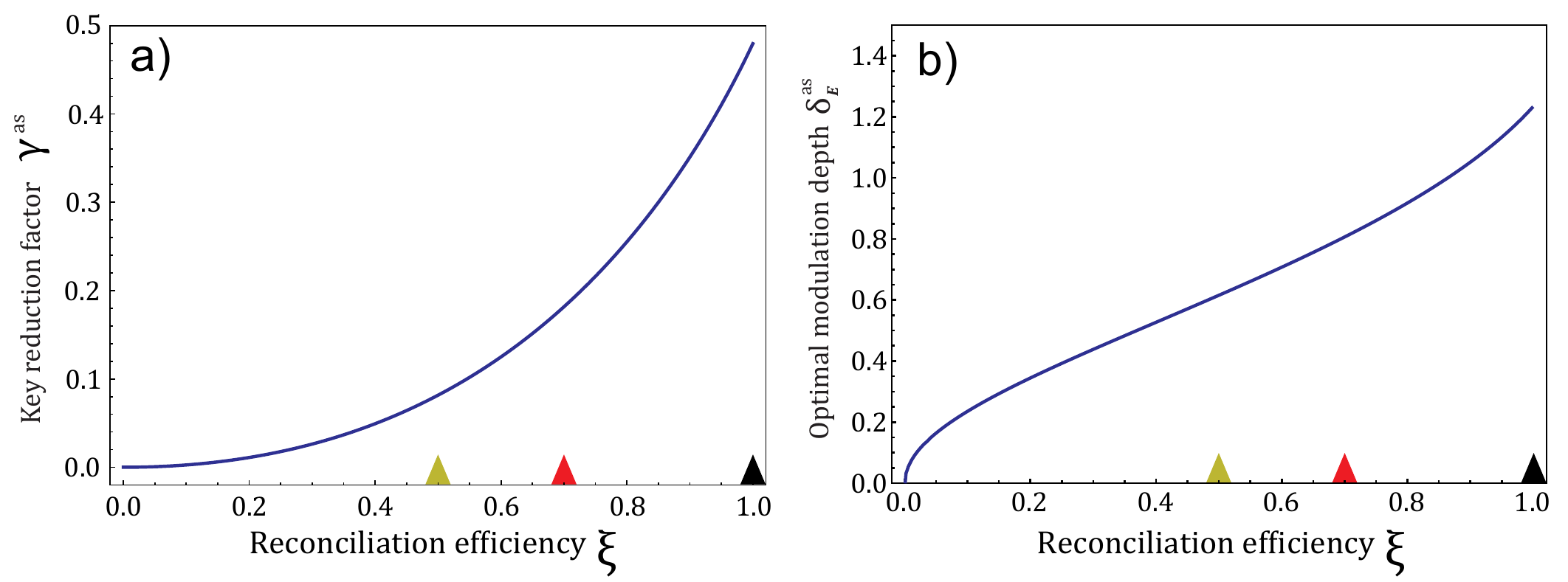}
\caption{(a) Reduction in the key amount $\gamma^{\textrm{as}}(\xi)$ in the asymptotic limit ${\cal R} \rightarrow 0$ given by Eq.~(\ref{Eq:gamma(beta)def}) as a function of the reconciliation efficiency $\xi$. (b) The corresponding modulation depth $\delta_E^{\textrm{as}}(\xi)$ in units determined by the normalized outcome of Eve's detection. Reconciliation efficiency values chosen for numerical examples depicted in Fig.~\ref{Fig:Key} are indicated with color arrows on the horizontal axes of the panels.}
\label{Fig:AsympSoft}
\end{figure}

In the soft decoding approach, Bob aims to infer the value of $q_A$ from the continuous outcome $y_B$. It is convenient to rewrite the expression (\ref{Eq:Keybeta}) for the key to the equivalent form
\begin{equation}
\label{Eq:KBsoft}
{\sf K}_{{\textrm{bin}},{\textrm{soft}}} = {\sf H}(B|E) - \xi {\sf H}(B|A) -(1-\xi) {\sf H}(B).
\end{equation}
In the above formula, the conditional entropy ${\sf H}(B|E)$ is calculated for the joint probability distribution
\begin{equation}
\label{Eq:p(y_B,y_E)}
p(y_B, y_E) = {\textstyle\frac{1}{4\pi}} \{ \exp[-(y_B+\delta_B)^2/2 -(y_E+\delta_E)^2/2]
+
\exp[-(y_B-\delta_B)^2/2 -(y_E-\delta_E)^2/2]\}.
\end{equation}
Further, the conditional entropy in the second term reads ${\sf H}(B|A) = \frac{1}{2}\log_2 (2\pi e)$, and the marginal entropy ${\sf H}(B)$ is given by \cite{KunzNJP2020}:
\begin{equation}
{\sf H}(B) = {\textstyle\frac{1}{2}}\log_2 (2\pi e) + \delta_B^2 \log_2 e
- \int_{-\infty}^{\infty} \frac{dt}{\sqrt{2\pi}} \exp[-(t-\delta_B)^2/2]
\log_2 [\cosh(\delta_B t)].
\end{equation}
The optimal modulation depth is found by maximizing ${\sf K}_{{\textrm{bin}},{\textrm{soft}}}$ given in Eq.~(\ref{Eq:KBsoft}) over $\delta_B$, or equivalently $\delta_E$, as these two variables are linearly dependent through Eq.~(\ref{Eq:deltaBdeltaErel}). Fig.~\ref{Fig:Key}(a) depicts the optimized key ${\sf K}^\ast_{{\textrm{bin}},{\textrm{soft}}}$ as a function of ${\cal R}$ for perfect reconciliation, $\xi=100\%$, as well as limited reconciliation efficiencies $\xi=70\%$ and $\xi=50\%$. The corresponding optimal argument values $\delta_E^\ast$ are shown in Fig.~\ref{Fig:Key}(b). It is seen that in the strong eavesdropping regime, when ${\cal R} \ll 1$, the key exhibits linear scaling with ${\cal R}$ and the parameter $\delta_E^\ast$ tends to a constant value dependent on the reconciliation efficiency $\xi$. It will be convenient to write
\begin{equation}
{\sf K}_{{\textrm{bin}},{\textrm{soft}}}^\ast \approx \gamma^{\textrm{as}}(\xi)  \times
{\textstyle \frac{1}{2}} {\cal R} \log_2 e
\label{Eq:K(binsoft)approx}
\end{equation}
where the multiplicative factor $\gamma^{\textrm{as}}(\xi)$ characterizes reduction in the key amount compared to the Gaussian upper bound in the asymptotic limit ${\cal R} \rightarrow 0$. As shown in Appendix~\ref{App:Soft}, the factor $\gamma^{\textrm{as}}(\xi)$ can be found by optimizing the expression
\begin{equation}
\label{Eq:gamma(beta)def}
\gamma^{\textrm{as}}(\xi) = \max_{\delta_E \ge 0}
\left\{ \delta_E^2 \left(1+\xi - \int_{-\infty}^{\infty} \frac{dt}{\sqrt{2\pi}} \,
\frac{\sinh(2\delta_E t)+1}{\cosh^2(\delta_E t)}
\exp[-(t - \delta_E)^2/2]\right)\right\}
\end{equation}
and the argument $\delta_E^{\textrm{as}}(\xi)$, for which the maximum on the right-hand side of the above expression is attained, defines the optimal modulation depth as detected by Eve. The asymptotic values match well the results of numerical optimization as seen in Fig.~\ref{Fig:Key}(b,c). The analysis of the strong eavesdropping scenario with soft decoding is summarized with Fig.~\ref{Fig:AsympSoft} which depicts the reduction in the key amount $\gamma^{\textrm{as}}(\xi)$ and the optimal modulation depth $\delta_E^{\textrm{as}}(\xi)$ as functions of the reconciliation efficiency $\xi$.

\subsection{Hard decoding}

The numerical complexity of key reconciliation can be reduced substantially by applying dual-threshold discrimination to Bob's detection outcome $y_B$ as described in Sec.~\ref{Sec:Model}. Such hard decoding produces a discrete variable
\begin{equation}
q_B = \begin{cases} 0 & \text{if $y_B < - \kappa$} \\
{\sf X} & \text{if $ -\kappa \le y_B \le \kappa$} \\
1 & \text{if $ y_B > \kappa$}
\end{cases}
\end{equation}
where $\kappa$ defines the discrimination threshold. Events ${\sf X}$ are treated as erasures and removed from further processing in the sifting step. The probability of generating a raw key bit is consequently given by
\begin{equation}
\label{Eq:praw}
p_{\textrm{raw}} = {\textstyle \frac{1}{2}} \bigl[ \textrm{erfc} \bigl( ({\kappa-\delta_B})/{\sqrt{2}}\bigr)
+ \textrm{erfc}\bigl( ({\kappa+\delta_B})/{\sqrt{2}}\bigr)
\bigr].
\end{equation}
The probability of error in the raw key is symmetric and reads
\begin{equation}
\label{Eq:proberror}
\varepsilon = (2p_{\textrm{raw}})^{-1} \textrm{erfc}\bigl( ({\kappa+\delta_B})/{\sqrt{2}}\bigr).
\end{equation}
%In the above expressions, $\sqrt{\cal R}\delta_E$ has been used in lieu of $\delta_B$ to facilitate optimization.
Under reverse reconciliation investigated here, the key amount per slot is given by
\begin{equation}
\label{Eq:Kbinhard}
{\sf K}_{{\textrm{bin}},{\textrm{hard}}} = p_{\textrm{raw}}\{\xi[1-{\sf H}_{\textrm{bin}}(\varepsilon)] - {\sf G}(\delta_E,\varepsilon)\},
\end{equation}
where $\xi$ is the reconciliation efficiency,
${\sf H}_{\textrm{bin}}(\varepsilon) = - \varepsilon \log_2 \varepsilon
- (1-\varepsilon) \log_2 (1-\varepsilon)$ is the binary entropy and the function ${\sf G}(\cdot,\cdot)$ is defined as:
\begin{multline}
{\sf G}(s,\varepsilon) = \int_{-\infty}^{\infty} \frac{dt}{\sqrt{2\pi}} \exp[-(t-s)^2/2]
\{ \varepsilon \log_2 [\varepsilon e^{st} + (1-\varepsilon) e^{-st}]  \\
+ (1-\varepsilon)\log_2 [\varepsilon e^{-st} + (1-\varepsilon) e^{st}]
- \log_2[\cosh(st)]\} .
\end{multline}
In order to optimize the key amount, the right hand side of Eq.~(\ref{Eq:Kbinhard}) needs to be maximized over the modulation depth characterized by $\delta_E$ and the discrimination threshold $\kappa$. Note that when $\delta_E$ is chosen as the free parameter in the optimization procedure, one needs to express $\delta_B = \sqrt{\cal R}\delta_E$ in Eqs.~(\ref{Eq:praw}) and (\ref{Eq:proberror}).

\begin{figure}
\centering
\includegraphics[width=1\columnwidth]{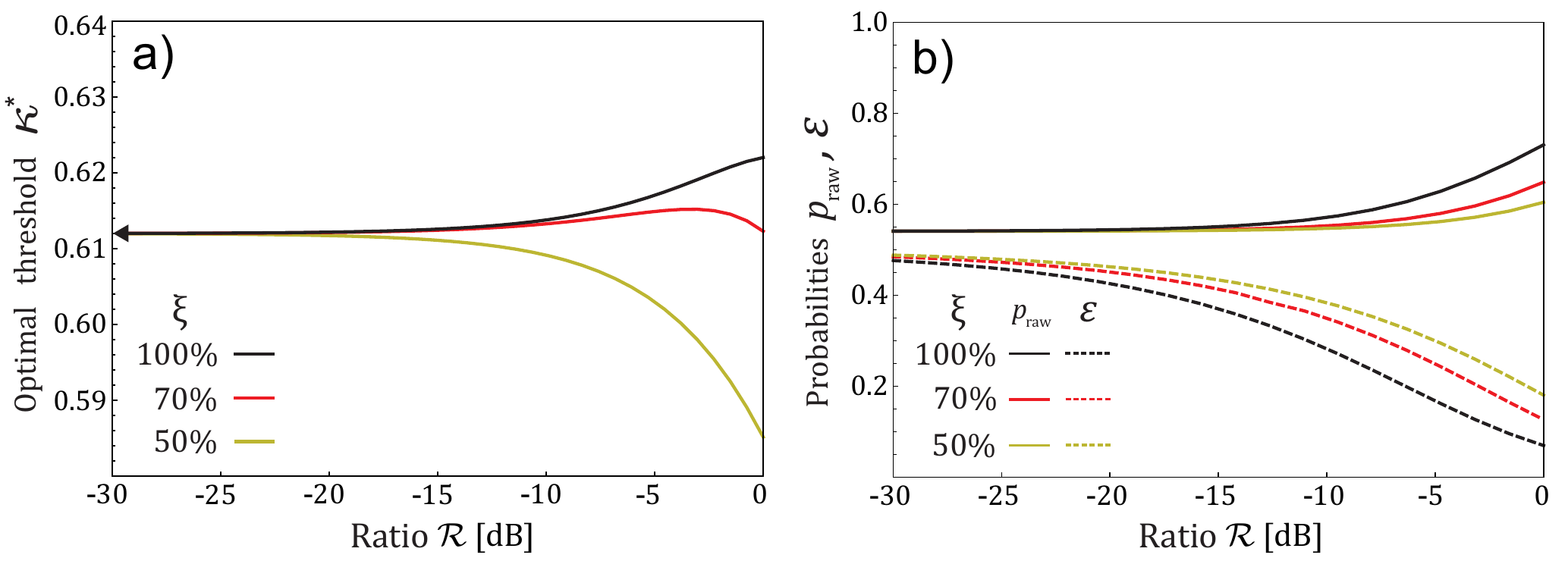}
\caption{(a) The optimal discrimination threshold $\kappa^\ast$ for the hard-decoded binary modulation scenario as a function of the ratio ${\cal R}$, shown for reconciliation efficiencies $\xi=100\%$ (black), $\xi=70\%$ (red) and $\xi=50\%$ (gold). The asymptotic value $\kappa^{\textrm{as}} = 0.6120$ derived in Appendix~\ref{App:Hard} is indicated with an arrow. (b) The resulting probability of generating a raw key bit $p_{\textrm{raw}}$ (solid lines) and the probability of error $\varepsilon$ (dashed lines) for the same selection of reconciliation efficiencies.}
\label{Fig:Hard}
\end{figure}

The optimized key ${\sf K}^\ast_{{\textrm{bin}},{\textrm{hard}}}$ and the corresponding optimal modulation depth $\delta_E^\ast$ as functions of ${\cal R}$ are shown respectively in Fig.~\ref{Fig:Key}(a) and (b) for reconciliation efficiencies $\xi=100\%, 70\%$, and $50\%$. It is seen that in the strong eavesdropping regime the key amount is reduced by a constant factor compared to the soft decoding case, while the optimal modulation depth remains to a very good approximation at the same level. Interestingly, the optimal discrimination threshold $\kappa^\ast$ weakly depends on either ${\cal R}$ or $\xi$, as seen in Fig.~\ref{Fig:Hard}(a). The resulting probability of generating a raw key bit $p_{\textrm{raw}}$ and the probability of error $\varepsilon$ are depicted in Fig.~\ref{Fig:Hard}(b).

The features identified above in numerical results can be characterized quantitatively with the help of the analysis of the asymptotic limit ${\cal R} \rightarrow 0$ presented in Appendix~\ref{App:Hard}. In this limit, the reduction in the key amount is given by a numerical factor
\begin{equation}
\label{Eq:Keybinhardapprox}
{\sf K}_{{\textrm{bin}},{\textrm{hard}}}^\ast \approx 0.8098
\times \gamma^{\textrm{as}} (\xi)  \times
{\textstyle \frac{1}{2}} {\cal R} \log_2 e.
\end{equation}
Interestingly, the factor $\gamma^{\textrm{as}} (\xi)$ as well as the optimal modulation depth $\delta^{\textrm{as}}_E (\xi)$ are the same as in the soft decoding case.
Furthermore, the optimal discrimination threshold $\kappa^{\textrm{as}} = 0.6120$ does not depend on the reconciliation efficiency.
The asymptotic values for the optimal modulation depth, the reduction of the key amount, and the optimal threshold are indicated with arrows on the vertical axis respectively in Figs.~\ref{Fig:Key}(b), \ref{Fig:Key}(c), and \ref{Fig:Hard}(a). It is seen that as long as the reconciliation efficiency is above $50\%$, the asymptotic values can be used in the analysis for the ratio ${\cal R}$ below approximately $-15$~dB.

\section{Discussion and conclusions}
\label{Sec:Conclusions}

The purpose of this work was to identify theoretically the optimal operation of an optical key distribution link utilizing weak intensity modulation, when an adversary implements a passive eavesdropping attack. In this scenario, the key security is ensured by the shot noise present in the signal detected by the eavesdropper. The key amount depends primarily on the ratio ${\cal R}$ defined in Eq.~(\ref{Eq:Rdef}) which compares the capabilities of the legitimate user and the eavesdropper to detect the optical signal. For Gaussian modulation, an upper bound on the attainable key is given by Eq.~(\ref{Eq:KG}). In the strong eavesdropping regime, when ${\cal R} \ll 1$ the key amount is simplified to a linear expression $\frac{1}{2}{\cal R}\log_2 e$. For other signal modulation formats and receivers setups considered in this work, this expression is reduced by a certain multiplicative factor.

In the case of binary modulation and soft decoding with reconciliation efficiency $\xi$, the key amount is reduced by a factor $\gamma^{\textrm{as}}(\xi)$ that can be calculated using Eq.~(\ref{Eq:gamma(beta)def}). The dependence of $\gamma^{\textrm{as}}(\xi)$  on $\xi$ is depicted in Fig.~\ref{Fig:AsympSoft}(a). The argument $\delta_E^{\textrm{as}}(\xi)$ maximizing the expression on the right hand side of Eq.~(\ref{Eq:gamma(beta)def}) specifies the optimal modulation depth: namely, the sender should use in individual slots optical energies given by $\bar{n} \pm \delta_E^{\textrm{as}}(\xi) \sigma_E/\tau_E$. Importantly, when ${\cal R} \ll 1$ the optimal modulation depth depends only on Eve's channel transmission $\tau_E$ and the detection noise $\sigma_E$ of her receiver, rather then corresponding parameters for Bob's link.

When dual-threshold hard decoding is used in the receiver, the key amount is further reduced by an approximately constant factor specified in Eq.~(\ref{Eq:Keybinhardapprox}). The optimal discrimination threshold exhibits weak dependence on either ${\cal R}$ or $\xi$. A possible difficulty in practical implementation of the binary modulation scheme with dual-threshold discrimination is the high probability of error in the raw key, as seen in Fig.~\ref{Fig:Hard}(b), which may limit the practically attainable reconciliation efficiency
\cite{ElkoussQIC2011}. Finally let us note that if there is uncertainty regarding Eve's eavesdropping capability, the worst-case scenario should be assumed. This will overestimate Eve's information about Bob's detection outcomes and hence ensure the security of the generated key, although may diminish its amount.

Qualitatively, the key per slot for all variants of the OKD protocol considered here has the same linear dependence on the ratio ${\cal R}$.
While Gaussian modulation offers the highest efficiency in terms of attainable key rate,
its practical implementation may be technically challenging and pose difficulties to attain high reconciliation efficiencies \cite{PRA99022322,Jouguet2011}. Therefore one can envisage that in practice the variant of choice will be binary modulation combined with hard-decision detection, whose overall simplicity should compensate for the minor reduction in the secret key rate.

\begin{table}[t]
\begin{center}
\begin{tabular}{ |l|l| }
  \hline
  \multicolumn{2}{|c|}{\textbf{LEO satellite orbit}} \\
  \hline
  Inclination angle & $51.64^{\circ}$  \\
  Orbit apogee/perigee & $420~\mathrm{km}$ \\
  Orbital period & 91 min 43 s \\
     \hline
  \multicolumn{2}{|c|}{\textbf{Transmitter parameters}} \\
    \hline
  Beam power & $1~\mathrm{W}$ \\
  Slot duration & $1~\mathrm{ns}$ \\
  Beam divergence (full angle) & $0.06^{\circ}$ \\
  Wavelength & $1550~\mathrm{nm}$ \\
   \hline
 \multicolumn{2}{|c|}{\textbf{Receiver parameters}} \\
   \hline
  Ground station latitude & $52^{\circ}\mathrm{N}$  \\
  Receiver telescope diameter & $40~\mathrm{cm}$ \\
  Detection power threshold & $-38~\mathrm{dBm}$ \\
  \hline
\end{tabular}
\end{center}
\caption{Parameters of a LEO-to-ground optical link used to estimate the length of a secure key that could be generated using the OKD technique during a single satellite pass.}
\label{Tab:Link}
\end{table}

To illustrate the practical potential of OKD we calculated the length of a secure key that can be generated during a single pass of a low Earth orbit (LEO) satellite equipped with an optical transmitter over an optical ground station (OGS) using parameters collected in Table~\ref{Tab:Link}. The parameter ${\cal R}$ has been taken as $-20$~dB, which describes a scenario where both Bob's and Eve's detectors are shot-noise limited, but Eve's receive telescope has ten times larger diameter than that of Bob. An equivalent scenario is equal apertures of Bob's and Eve's telescopes, but Bob's detector noise approx.\ $20$~dB above the shot noise level. For a pass with a maximum elevation angle of $55^{\circ}$ the optical power received by the OGS exceeds the detection threshold during an interval of $122~\mathrm{s}$. Taking the slot rate of 1~Gbaud/s and theoretical limit for binary modulation and hard-decoding with 100\% reconciliation efficiency yields the total length of the generated key equal to 340~Mbits corresponding to the key rate of 2.79~Mbps. For a non-unit reconciliation efficiency $\xi$ these figures are reduced by a factor depicted in Fig.~\ref{Fig:AsympSoft}(a).
The presented example indicates that for passive eavesdropping attacks OKD can provide secure key rates contending with those offered by conventional QKD \cite{Pan2020} while dispensing with some of elaborate technologies required by the latter.

\section*{Acknowledgments}
We wish to acknowledge insightful discussions with
K. Inoue, P. V. Trinh, J. Ko{\l}ody\'{n}ski, M. Parniak-Niedojad{\l}o, and W. Wasilewski.

\section*{Funding}
This work is a part of the project ``Quantum Optical Technologies'' carried out within the International Research Agendas programme of the Foundation for Polish Science co-financed by the European Union under the European Regional
Development Fund. ML and PK acknowledge financial support by the Foundation for Polish Science (FNP) (project First Team co-financed by the European Union under the European Regional Development Fund, POIR.04.04.00-00-3FD9/17).

\section*{Data availability}
Data underlying the results presented in this paper are not publicly available at this time but may be obtained from the authors upon reasonable request.

\section*{Disclosures}
The authors declare no conflicts of interest.

\appendix

\section{Asymptotics for binary modulation, soft decoding}
\label{App:Soft}

The conditional entropy ${\sf H}(B|E)$ entering Eq.~(\ref{Eq:KBsoft}) can be expressed as
\begin{equation}
{\sf H}(B|E) = - \int_{-\infty}^{\infty} dy_B\int_{-\infty}^{\infty} dy_E
p(y_B,y_E) \log_2 p(y_B|y_E)
\end{equation}
which involves the joint probability distribution $p(y_B,y_E)$ specified in Eq.~(\ref{Eq:p(y_B,y_E)}) and the
conditional probability distribution $p(y_B|y_E)$ that can be written as
\begin{equation}
p(y_B|y_E) = \frac{p(y_B,y_E)}{\int_{-\infty}^{\infty} dy_B \, p(y_B, y_E)} =
\frac{\cosh(\delta_E y_E + \delta_B y_B)}%
{\sqrt{2\pi}\cosh(\delta_E y_E )}
\exp(-y_B^2/2 - \delta_B^2/2).
\end{equation}
Consequently, the expression for the conditional entropy ${\sf H}(B|E)$ can be simplified to the form
\begin{equation}
\label{Eq:H(B|E)simplified}
{\sf H}(B|E) = {\textstyle\frac{1}{2}} \log_2 (2\pi e) + \delta_B^2 \log_2 e
- \int_{-\infty}^{\infty} dy_B \int_{-\infty}^{\infty} dy_E \, p(y_B, y_E)
 \log_2 \frac{\cosh(\delta_E y_E + \delta_B y_B)}{\cosh(\delta_E y_E )}.
\end{equation}
%Consequently, the expression for the key ${\sf K}_{{\textrm{bin}},{\textrm{soft}}}$ can be simplified to the form [Groch z kapusta sie zrobil, bo tu juz jest przyblizenie $H(B)$].
%\begin{equation}
%\label{Eq:KBsoftsimplified}
%{\sf K}_{{\textrm{bin}},{\textrm{soft}}} = {\textstyle\frac{1}{2}}(1+\xi)\delta_B^2 \log_2 e
%- \int_{-\infty}^{\infty} dy_B \int_{-\infty}^{\infty} dy_E \, p(y_B, y_E)
% \log_2 \frac{\cosh(\delta_E y_E + \delta_B y_B)}%
%{\cosh(\delta_E y_E )}.
%\end{equation}
Numerical optimization indicates that for strong eavesdropping,when ${\cal R} \ll 1$ the optimal value $\delta_E^\ast$ is of the order of one. Hence in this regime $\delta_B =\sqrt{\cal R} \delta_E $ becomes small compared to one. This justifies expanding the logarithm in Eq.~(\ref{Eq:H(B|E)simplified}) up to the quadratic term in $\delta_B$, which yields
\begin{equation}
\log_2 \frac{\cosh(\delta_E y_E + \delta_B y_B)}%
{\cosh(\delta_E y_E )} \\
\approx \left( \delta_B y_B \tanh(\delta_E y_E) +
\frac{(\delta_B y_B)^2}{2\cosh^2(\delta_E y_E)}\right) \log_2 e
\end{equation}
and allows one to perform integration over $y_B$.
Further, the marginal entropy ${\sf H}(B)$ can be approximated by \cite{KunzNJP2020}
\begin{equation}
{\sf H}(B) \approx {\textstyle\frac{1}{2}} \log_2 (2\pi e) + {\textstyle\frac{1}{2}}
\delta_B^2 \log_2 e.
\end{equation}
This leaves one with the following approximate expression for the key in the leading order of $\delta_B^2 = {\cal R} \delta_E^2$:
\begin{equation}
{\sf K}_{{\textrm{bin}},{\textrm{soft}}} \approx
\delta_E^2 \left(1+\xi - \int_{-\infty}^{\infty} \frac{dt}{\sqrt{2\pi}} \,
%\right. \\ \times \left. \vphantom{\int}
\frac{\sinh(2\delta_E t)+1}{\cosh^2(\delta_E t)}\exp[-(t - \delta_E)^2/2] \right) \times
{\textstyle \frac{1}{2}}
{\cal R}   \log_2 e
\end{equation}
where for the sake of clarity the integration variable has been changed from $y_E$ to $t$.
In order to identify the optimal modulation depth for a given reconciliation efficiency $\xi$, the factor $\delta_E^2 (\ldots)$ needs to be maximized over $\delta_E$, as specified in Eq.~(\ref{Eq:gamma(beta)def}).

\section{Hard decoding in the asymptotic limit}
\label{App:Hard}

Analogously to the soft decoding scenario, numerics suggests that for ${\cal R} \ll 1$ one should consider $\delta_B \ll 1$. Expansion of the probability of generating a raw key bit $p_{\textrm{raw}}$ and the probability of error $\varepsilon$ up to the second order in $\delta_B$ yields:
\begin{equation}
p_{\textrm{raw}}  = \textrm{erfc}(\kappa/\sqrt{2}) + O(\delta_B^2) , \qquad
\varepsilon  = \frac{1}{2} \left( 1- \sqrt{\frac{2}{\pi}}
\frac{\exp(-\kappa^2/2)}{\textrm{erfc}(\kappa/\sqrt{2})} \delta_B  \right) + O(\delta_B^2).
\end{equation}
Given the above approximate form of $\varepsilon$,
in the following it will be convenient take $\varepsilon = (1-\eta)/2$ and expand ${\sf G}\bigl(s, (1-\eta)/2\bigr) $ around $\eta=0$:
\begin{equation}
{\sf G}\bigl(s, (1-\eta)/2\bigr) \approx \eta^2 \log_2 e
\times \int_{-\infty}^{\infty} \frac{dt}{\sqrt{2\pi}}
 [\tanh(st) - {\textstyle \frac{1}{2}} \tanh^2(st)] \exp[-(t-s)^2/2].
\end{equation}
Note that
\begin{equation}
\tanh(st) - {\textstyle \frac{1}{2}} \tanh^2(st) =
\frac{\sinh(2st) +1}{2\cosh^2(st)} - \frac{1}{2}.
\end{equation}
Further, for $\eta \ll 1$ mutual information for a binary symmetric channel can be approximated by
\begin{equation}
1- {\sf H}_{\textrm{bin}} \bigl( (1-\eta)/2 \bigr) \approx {\textstyle \frac{1}{2}} \eta^2 \log_2 e.
\end{equation}

Inserting the approximate expressions into Eq.~(\ref{Eq:Kbinhard}) yields in the leading order quadratic in $\delta_B^2 = {\cal R}\delta_E^2$:
\begin{multline}
K_{{\textrm{bin}},{\textrm{hard}}} \approx \frac{2}{\pi}\frac{\exp(-\kappa^2)}{\textrm{erfc}(\kappa/\sqrt{2})}
\times \delta_E^2 \left( 1+ \xi - \int_{-\infty}^{\infty} \frac{dt}{\sqrt{2\pi}}
\frac{\sinh(2\delta_E t)+1}{\cosh^2(\delta_E t)} e^{-(t-\delta_E)^2/2} \right)
\times {\textstyle \frac{1}{2}}{\cal R} \log_2 e
\end{multline}
Interestingly, the discrimination threshold $\kappa$ and the modulation depth $\delta_E$ enter two separate multiplicative factors. Hence, optimization with respect to these two parameters can be carried out independently. Moreover, the factor $2 \exp(-\kappa^2)/[\pi \textrm{erfc}(\kappa/\sqrt{2})]$ dependent on $\kappa$ does not involve the reconciliation efficiency $\xi$. Its numerical optimization yields the maximum value equal to $0.809826$ that is obtained for the argument $\kappa^{\textrm{as}} = 0.612003$.  The second factor, dependent on $\delta_E$ has exactly the same form as in the case of soft decoding and its optimization yields again the factor $\gamma^{\textrm{as}}(\xi)$ given by Eq.~(\ref{Eq:gamma(beta)def}).

\bibliography{imddokd}

\end{document}